\documentclass[aps,prl,showpacs,twocolumn,floats]{revtex4}

\usepackage{graphicx,psfig}
\usepackage{dcolumn}
\usepackage{bm}

\begin{document}

\bibliographystyle{prsty}

\title{
%
Universal decoherence in solids \vspace{-1mm} }
\author{
Eugene M. Chudnovsky}
 \affiliation{\mbox{Department of Physics and Astronomy,} \\
\mbox{Lehman College, City University of New York,} \\ \mbox{250
Bedford Park Boulevard West, Bronx, New York 10468-1589, U.S.A.}
\\
 {\rm (Received 3 December 2003)}
}

\begin{abstract}
Symmetry implications for the decoherence of quantum oscillations
of a two-state system in a solid are studied. When the oscillation
frequency is small compared to the Debye frequency, the universal
lower bound on the decoherence due to the atomic environment is
derived in terms of the macroscopic parameters of the solid, with
no unknown interaction constants.
\end{abstract}


\maketitle

The problem of tunneling and coherence in the presence of
dissipation has been extensively studied in the past, see, e.g.,
the review of Leggett et al. \cite{Leggett-review}. In this Letter
we shall revise the case of quantum oscillations of a particle (an
electron, an atom or a defect) in a double-well potential in a
solid in the presence of weak dissipation due to the
non-conducting atomic environment. We shall demonstrate that all
previous works on this subject have missed two important facts.
The first fact is that the double-well potential formed by the
local arrangement of atoms in a solid is defined in the coordinate
frame of that local atomic environment, not in the laboratory
frame. The second fact is that the interactions of the tunneling
variable with phonons must be invariant with respect to global
translations and rotations. When these facts are taken into
account, a simple universal result for the decoherence rate can be
obtained.

Let the degenerate minima of the double-well potential be located
at the points $X_{0}$ and $-X_{0}$ inside the solid. In the case
of weak dissipation, if, at the moment of time $t=0$, the particle
is prepared in the state $X = X_{0}$, then for any moment $t > 0$
\begin{equation}
\langle X(0)X(t) \rangle = X_{0}^{2}e^{-{\Gamma}t}
\cos({\omega}_{0}t) \,,
\end{equation}
where $\langle ... \rangle$ means quantum average,
${\hbar}{\omega}_{0} = \Delta$ is the tunneling splitting, and
${\Gamma} \ll {\omega}_{0}$ is the decoherence rate. When
addressing the effect of dissipation one should answer two
questions: 1) How is ${\Delta}$ renormalized by the atomic
environment? and 2) What is the value of $\Gamma$? The answer to
the first question requires the precise knowledge of the
interactions with the environmental degrees of freedom. After the
latter are integrated out from the total action \cite{Feynman},
$\Delta$ can be computed via the instanton of the remaining
effective action for the tunneling variable \cite{Sethna,CL}. For
the ohmic interactions, $\Delta$ can be nicely expressed in terms
of a single measurable dissipation constant \cite{CL}. In this
Letter we show that a similar beautiful result follows for the
decoherence due to the superohmic interactions with phonons. In
the limit when ${\omega}_0$ is small compared to the Debye
frequency, ${\omega}_D$, the decoherence rate $\Gamma$ can be
expressed in terms of measurable constants of the solid, with no
unknown interaction constants. Similar results for problems
involving tunneling of the angular momentum have been obtained
earlier \cite{CM,CK}. Here we present rigorous derivation of the
universal expression for $\Gamma$ due to phonons for an arbitrary
double-well potential in a solid.

The Hamiltonian of the system is
\begin{eqnarray}\label{Hamiltonian}
{\cal{H}} & = &  \frac{1}{2}\, {\rm m}{\dot{\bf R'}}^2 + U({\bf R})
+ {\alpha}_{iklm}R_{i}R_{k}u_{lm}({\bf R}) + ... \nonumber \\
& + & \int d^{3}r \left(\frac{1}{2}\rho {\dot{\bf u}}^{2} +
{\lambda}_{iklm}u_{ik}u_{lm}\right) \, ,
\end{eqnarray}
where, ${\bf R'}$ and ${\bf R}$ are the radius-vectors of the
particle of mass ${\rm m}$ in the laboratory coordinate frame and
in the coordinate frame rigidly coupled with a solid respectively,
$U({\bf R})$ is the double-well potential, ${\bf u}({\bf r})$ is
the phonon displacement field, $u_{ij}=\frac{1}{2}({\partial}_{i}
u_{j}+{\partial}_{j}u_{i})$ is the strain tensor, $\rho$ is the
mass density of the solid, ${\lambda}_{iklm}$ is the tensor of
elastic coefficients, and ${\alpha}_{iklm}, ...$ are coefficients
of the expansion of the long-wave interaction between ${\bf R}$
and ${\bf u}$ into a series on $u_{ij}$. The invariance of the
interaction with respect to the translation [${\bf u}({\bf r})
\rightarrow {\bf u}({\bf r}) + {\bf a}$] and rotation [${\bf
u}({\bf r}) = {\bf \Phi} \times {\bf r}$] of a solid as a whole
rules out combinations of the form ${\bf R}{\cdot}{\bf u}$ and
${\bf R}{\cdot}{\bm \nabla}{\times}{\bf u}$. All interaction
terms, therefore, must be even in the number of the spatial
components of ${\bf R}$. Our result is based, in part, upon this
observation. Another important observation is that the kinetic
energy of the particle depends on ${\dot{\bf R'}}$, while the
potential energy depends on ${\bf R}$. Substituting ${\bf R'} =
{\bf R} + {\bf u}$ into Eq.\ (\ref{Hamiltonian}), one obtains
\begin{eqnarray}\label{Hamiltonian-R}
{\cal{H}} & = &  \frac{1}{2}\, {\rm m}{\dot{\bf R}}^2 + U({\bf R})
\nonumber \\
& + & {\rm m}{\dot{\bf R}}{\cdot}\dot{\bf u}({\bf R})  +
{\alpha}_{iklm}R_{i}R_{k}u_{lm}({\bf R}) + ... \nonumber \\
& + & \int d^{3}r \left(\frac{1}{2}[\rho + {\rm m}{\delta}({\bf r}
- {\bf R} )] {\dot{\bf u}}^{2} +
{\lambda}_{iklm}u_{ik}u_{lm}\right) \, .
\end{eqnarray}
The renormalization of the mass density in Eq.\
(\ref{Hamiltonian-R}) is totally insignificant in the long-wave
limit (see below), but the presence of the effective dynamic
interaction, ${\rm m}{\dot{\bf R}}{\cdot}\dot{\bf u}$, which is
linear in the tunneling variable, is absolutely crucial for our
argument.

In this Letter we will treat the renormalized value of ${\Delta}$
as a known parameter that can be obtained from experiment. We
shall start with a symmetric double-well potential. Let the two
degenerate minima of $U({\bf R})$ be ${\bf R}=\pm{\bf R}_{0}$, so
that the ground state and the first excited state of the particle
are
\begin{eqnarray}\label{states}
|0 \rangle & = & \frac{1}{\sqrt{2}}(|{\bf R}_{0} \rangle +
|-{\bf R}_{0} \rangle) \nonumber \\
|1 \rangle & = & \frac{1}{\sqrt{2}}(|{\bf R}_{0} \rangle - |-{\bf
R}_{0} \rangle)\;.
\end{eqnarray}
We shall assume that the energy gap, $\Delta$, between these two
states, renormalized by the environment, is small compared to the
distance to other energy levels of the particle. Our goal is to
compute the decoherence of the low-energy states, $|\psi \,\rangle
= C_1 \, |0 \rangle + C_2 \, |1 \rangle$, due to the decay of $|1
\rangle$ into $|0 \rangle$, accompanied by the emission of a
phonon of frequency ${\omega}_{0} = {\Delta}/{\hbar}$. In the
limit when ${\omega}_{0} \ll {\omega}_{D}$, the phonon wavelength
is large compared to the interatomic distance, which justifies the
use of the long-wave limit of the elastic theory. To simplify
calculations, we shall also assume that the tunneling distance,
$2X_{0}$, is small compared to the wavelength of phonons of
frequency ${\omega}_{0}$. Writing ${\bf R}_{0}$ as $(X_{0}, 0 ,
0)$, one has $\hat{X} \, |\pm {\bf R}_{0} \rangle = \pm X_{0} \,
|\pm {\bf R}_{0} \rangle $, while $\hat{Y} \, |\pm {\bf R}_{0}
\rangle = \hat{Z} \, |\pm {\bf R}_{0} \rangle = 0$. Consequently,
$\hat{X} \, |0 \rangle = X_{0} \, |1 \rangle $ and $\hat{X} \, |1
\rangle = X_{0} \, |0 \rangle $, while $\hat{Y}$ and $\hat{Z}$
produce zero result when acting on $|0 \rangle$ and $|1 \rangle$.
It is now easy to see that any combination of the even number of
operators $\hat{X}$, $\hat{Y}$, and $\hat{Z}$, including
${\hat{X}}^{2}$, etc., has a zero matrix element between $|0
\rangle$ and $|1 \rangle$. Thus, the only decohering term in Eq.\
(\ref{Hamiltonian-R}) is ${\rm
m}\hat{\dot{X}}{\hat{\dot{u}}}_{x}$, with
\begin{equation}\label{X}
\langle 0|\hat{\dot{X}}|1 \rangle  = \langle
0|\frac{i}{\hbar}(\hat{\cal{H}}\hat{X} - \hat{X}\hat{\cal{H}})|1
\rangle = - i{\omega}_{0}X_0 \,,
\end{equation}
where it is explicitly implied that $|0 \rangle$ and $|1 \rangle$
are the approximate eigenstates of the full Hamiltonian, so that
$\Delta = \langle 1 | \hat{\cal{H}} |1 \rangle - \langle 0 |
\hat{\cal{H}} |0 \rangle$ is the tunneling splitting renormalized
by the interactions.

We first consider the case of zero temperature. The $T=0$
decoherence rate is given by the Fermi golden rule:
\begin{eqnarray}\label{Fermi}
& \Gamma & =  \frac{2\pi}{\hbar} \sum_{i\neq j} <i|{\rm
m}\hat{{\dot{\bf R}}}{\cdot}\hat{\dot{\bf u}}|j>
  <j|{\rm m}\hat{\dot{\bf R}}{\cdot}\hat{\dot{\bf u}}|i> \delta(E_i - E_j) \nonumber \\
  & = & \frac{2\pi}{\hbar}{\rm m}^{2}\, |\langle 0 | \hat{\dot{X}}
  | 1 \rangle |^{2}\sum_{{\bf k},i} |\langle {\bf k},i |
  {\hat{{\dot{u}}}}_{x}| 0_{ph} \rangle |^{2} \delta(\hbar
  {\omega}_{{\bf k}i} - \Delta) \,
\end{eqnarray}
where $| 0_{ph} \rangle$ and $| {\bf k},i \rangle$ are the states
of the solid with zero phonons and one phonon of the wave vector
${\bf k}$ and polarization $i$ respectively, and ${\omega}_{{\bf
k}i} = c_i k$ is the phonon frequency, with $c_i$ being the speed
of sound of the polarization $i$. The canonical quantization of
the phonon field \cite{Kittel} yields
\begin{equation}\label{u}
{\hat{{\dot{u}}}}_{x} = \frac{-i}{\sqrt{V}} \sum_{{\bf
k},i}\sqrt{\frac{{\hbar}{\omega}_{{\bf k}i}}{2{\rho}}}\left(
a_{{\bf k}i}{\rm e}^{i\bf k r} - a^\dagger_{{\bf k}i}{\rm
e}^{-i\bf k r} \right)e^x_{i}\;,
\end{equation}
where $V$ is the volume of the system, $a^\dagger_{{\bf k}i}$ and
$a_{{\bf k}i}$ are operators of creation and annihilation of
phonons, and ${\bf e}_{i}$ are unit polarization vectors.
Substitution of Eq.\ (\ref{X}) and Eq.\ (\ref{u}) into Eq.\
(\ref{Fermi}), with account of $\langle (e^x_{i})^{2} \rangle =
1/3$, gives
\begin{equation}\label{sum}
\Gamma = \frac{\pi {\rm
m}^{2}X_{0}^{2}{\omega}_{0}^{2}}{3{\hbar}{\rho}V} \sum_{{\bf k},i}
{\omega}_{{\bf k}i} \, \delta({\omega}_{{\bf k}i} -
{\omega}_{0})\,.
\end{equation}
Replacing the summation over ${\bf k}$ by $V\int d^3k/(2 \pi)^{3}$
one obtains
\begin{equation}\label{T=0}
\Gamma = \frac{{\rm m}^{2}X_0^2{\omega}_{0}^{5}}{6 \pi {\hbar}
\rho} \left(\frac{2}{c_t^3} + \frac{1}{c_l^3}\right) \,,
\end{equation}
where $c_t$ and $c_l$ are the speeds of the transversal and
longitudinal sound respectively.

The following observation allows one to understand the above
result in simple physical terms. Up to a numerical factor of order
unity, $\Gamma$ of Eq.\ (\ref{T=0}) satisfies
\begin{equation}\label{momentum}
{\hbar}{\Gamma} \, \sim \, p_0^2/2M \, ,
\end{equation}
with $p_0 = mX_0{\omega}_0$ being the rms value of the momentum of
the oscillating particle and $M \, \sim \, {\rho} {\lambda}^3$
being the mass of the solid within the volume of dimensions
$\lambda$, where $\lambda$ is the wavelength of phonons of
frequency ${\omega}_0$, averaged over polarizations. This volume
of the solid adjacent to the particle must oscillate together with
the particle in order to conserve the linear momentum. The latter
is a consequence of the commonly neglected fact that the potential
$U({\bf R})$ is formed by atoms that are coupled to the rest of
the solid, which makes the double-well a part of the dissipative
environment. The $p_0^2/2M$ width of the excited state is,
therefore, a consequence of the conservation law that mandates the
entanglement of the particle states with the states of the solid.

The superohmic case has been studied in Ref.
\onlinecite{Leggett-review}, based upon the spin-boson
Hamiltonian,
\begin{eqnarray}\label{spin-boson}
{\cal{H}}_{SB}  & = & -{\Delta}_{0}{\hat{s}}_{x} +
X_{0}{\hat{s}}_{z} \sum_{\alpha} C_{\alpha}x_{\alpha} \nonumber
\\
& + & \sum_{\alpha} \left(\frac{p^{2}_{\alpha}}{2{\rm m}_{\alpha}}
+ \frac{1}{2}{\rm m}_{\alpha}{\omega}^{2}_{\alpha}x_{\alpha}^{2}
\right) \,.
\end{eqnarray}
Here $\hat{\bf s}$ is the spin-1/2 operator, ${\Delta}_{0}$ is the
bare splitting, ${\alpha}$ labels coordinates $x_{\alpha}$,
momenta $p_{\alpha}$, frequencies ${\omega}_{\alpha}$, and masses
${\rm m}_{\alpha}$ of harmonic oscillators, and $C_{\alpha}$ are
constants of the linear coupling between the macroscopic variable
$X$ and the oscillators. The following result was obtained for the
decoherence rate \cite{Leggett-review}: $\Gamma =
(X_{0}^{2}/4{\hbar})J(\Delta)$, where $J(\omega) = \sum_{\alpha}
(\pi C^{2}_{\alpha}/2{\rm m}_{\alpha}{\omega}_{\alpha})
\delta(\omega - {\omega}_{\alpha})$ and ${\Delta}$ is the
renormalized splitting. This, of course, is a correct answer to
the mathematical problem posed by Eq.\ (\ref{spin-boson}).
According to Ref. \onlinecite{Leggett-review}, for a nonlinear
coupling of the form $F_{\alpha}(X)x_{\alpha}$, the quantity
$C_{\alpha}$ in the above formulas should be replaced by
$X_{0}^{-1}[F_{\alpha}(\frac{1}{2}X_{0}) -
F_{\alpha}(-\frac{1}{2}X_{0})]$. Then, when only even powers of
$X$ in $F_{\alpha}(X)$ are allowed by symmetry, the result for the
decoherence rate is zero. As we have seen above, in a physical
problem of quantum oscillations of a particle or a defect in a
double-well potential in a solid the linear coupling of the
tunneling variable to the phonons is prohibited by symmetry,
unless one identifies the coupling with the Galilean
transformation term ${\rm m}{\dot{\bf R}}{\cdot}\dot{\bf u}$ and
chooses $C_{\alpha} \propto {\omega}_{\alpha}^{2}$. We, therefore,
conclude, that in relevant physical problems the only correct
answer for $\Gamma$ is given by Eq.\ (\ref{T=0}). The beauty of
this answer is that it does not depend on any unknown interaction
constants but is expressed in terms of measurable parameters only.

We shall now generalize our answer to include the case of an
asymmetric double-well and finite temperature. We shall start with
the first and will assume that the energy minima of $U({\bf R}$)
are shifted with respect to one another by a small energy
$\epsilon$. The origin of the coordinate system can always be
chosen such that these minima correspond to ${\bf R} = \pm {\bf
R}_{0}$, with ${\bf R}=(X_0, 0, 0)$. The relevant two-state
Hamiltonian of the particle is
\begin{equation}\label{twostate}
h   =  -{\Delta}{\hat{s}}_{x} + {\epsilon}{\hat{s}}_{z} \,.
\end{equation}
The solution of the corresponding Schr\"{o}dinger equation yields:
\begin{eqnarray}\label{epsilon-states}
|0 \rangle & = & \frac{1}{\sqrt{2}}(C_{-}|{\bf R}_{0} \rangle +
C_{+}|-{\bf R}_{0} \rangle) \nonumber \\
|1 \rangle & = & \frac{1}{\sqrt{2}}(C_{+}|{\bf R}_{0} \rangle -
C_{-}|-{\bf R}_{0} \rangle)\;,
\end{eqnarray}
where
\begin{equation}\label{C}
C_{\pm} = \sqrt{1 \pm {\epsilon}/\sqrt{{\Delta}^{2} +
{\epsilon}^{2}}} \,.
\end{equation}
The energy gap between these states is
\begin{equation}\label{gap}
{\hbar}{\omega} = \sqrt{{\Delta}^{2} + {\epsilon}^{2}} \,.
\end{equation}
It is easy to see that the states (\ref{epsilon-states}) are still
the eigenstates of ${\hat{X}}^{2}$, as for ${\epsilon}=0$, so that
$\langle 0|{{\hat{R}}}_{i}{{\hat{R}}}_{j}|1 \rangle = 0$, as
before. Among terms in Eq.\ (\ref{Hamiltonian-R}) which are linear
in $\hat{\bf R}$, the matrix element responsible for the
decoherence continues to be
\begin{equation}\label{X-epsilon}
\langle 0|\hat{\dot{X}}|1 \rangle  = - i{\omega}X_0 C_{+}C_{-} = -
i{\omega}_{0}X_0 \,,
\end{equation}
which is independent of ${\epsilon}$. The phonon part in Eq.\
(\ref{Fermi}), however, must be modified by replacing ${\Delta}$
with the full gap, ${\hbar}{\omega}=\sqrt{{\Delta}^{2} +
{\epsilon}^{2}}$, in the $\delta$-function. At a finite
temperature one should sum up the transitions between the initial
state, $|1 \rangle$, with $n_{{\bf k}i}$ phonons and the final
state, $|0 \rangle$, with $n_{{\bf k}i} + 1$ phonons, or vice
versa for the transition $| 0 \rangle \rightarrow |1 \rangle$.
This gives
\begin{equation}\label{sum-T}
\Gamma = \frac{\pi {\rm
m}^{2}X_{0}^{2}{\omega}_{0}^{2}}{3{\hbar}{\rho}V} \sum_{{\bf k},i}
{\omega}_{{\bf k}i} \,(2n_{{\bf k}i} + 1)\, \delta({\omega}_{{\bf
k}i} - {\omega})
\end{equation}
instead of Eq.\ (\ref{sum}). The integration over the phonon modes
must be accompanied by thermal averaging over the number of
phonons, with $\langle n_{{\bf k}i} \rangle = [\exp(\hbar
{\omega}_{{\bf k}i}/T) - 1]^{-1}$. The final answer for a biased
double-well at $T \neq 0$ reads:
\begin{equation}\label{T}
\Gamma = \frac{{\rm m}^{2}X_0^2{\omega}_{0}^{2}{\omega}^{3}}{6 \pi
{\hbar} \rho} \left(\frac{2}{c_t^3} +
\frac{1}{c_l^3}\right)\coth\left(\frac{{\hbar}{\omega}}{2T}\right)
\,.
\end{equation}
Once again we emphasize that it does not contain any unknown
interaction constants. The effect of the interactions has been
entirely absorbed into the renormalized splitting,
${\hbar}{\omega}_0$.

Among various tunneling problems, the problem studied above is
relevant to the width of a low-energy optical mode that
corresponds to quantum oscillations of an atom between two
isomeric positions in the unit cell of a crystal. It describes a
kind of an ammonia-molecule arrangement of atoms imbedded in a
solid. For, e.g., an atom of mass ${\rm m} \sim
3{\times}10^{-23}\,$g, oscillating at ${\omega}_{0} \sim
10^{12}\,$s$^{-1}$ in a symmetric double-well with $X_{0} \sim
2{\times}10^{-8}\,$cm in a crystal with ${\rho} \sim 5\,$g/cm$^3$
and $c_t \sim 10^5\,$cm/s $< c_l$ at $T < {\hbar}{\omega}_{0}$,
the width given by Eq.\ (\ref{T}) is of the order of
$10^{10}\,$s$^{-1}$. For tunneling of electrons, the effect is
generally small due to the smallness of the electron mass. One
should remember, however, that $\Gamma$ of Eq.\ (\ref{T})
represents the ultimate lower limit of the decoherence rate which
is mandated by the conservation law.

To have a complete picture we will show how the above treatment of
the particle problem can be transformed to include the problems of
the decoherence of quantum oscillations of the angular variable,
e.g. spin or an orbital moment. This problem is relevant to
tunneling of the magnetic moment of a molecule between up and down
directions in a crystal field \cite{book,CM}. Here we will give
its general solution in the presence of a bias field. When the
tunneling variable is the angular momentum ${\bf L}$, the
potential $U({\bf R})$ in Eq.\ (\ref{Hamiltonian-R}) must be
replaced with
\begin{equation}\label{Hamiltonian-L}
{\cal{H}}_{0} = {\beta}_{ik}L_{i}L_{k} +
{\beta}'_{iklm}L_iL_kL_lL_m + ... \,,
\end{equation}
where ${\beta}_{ik}, {\beta}'_{iklm}$, etc. are tensors determined
by the symmetry of the crystal. The analogy with the particle
problem is that ${\bf L}$ in Eq.\ (\ref{Hamiltonian-L}) is the
angular momentum in the coordinate frame coupled to the local
crystal axes.

Rotations of ${\bf L}$ with respect to the crystal axes couple to
the transversal phonon modes satisfying ${\bm \nabla}{\cdot}{\bf
u} = 0$. Consequently, $\dot{\bf u}({\bf R})$ can be written as
${\bf \Omega}{\times}{\bf R}$, with ${\bf \Omega}$ given by ${\bf
\Omega}=\frac{1}{2}{\bm \nabla}{\times}\dot{\bf u}$. This allows
one to write ${\rm m}{\dot{\bf R}}{\cdot}\dot{\bf u}({\bf R})$ in
Eq.\ (\ref{Hamiltonian-R}) in terms of the angular momentum,
\begin{equation}\label{L-term}
{\rm m}{\dot{\bf R}}{\cdot}\dot{\bf u}({\bf R}) = ({\bf
R}{\times}{\rm m}{\dot{\bf R}}){\cdot}{\bf \Omega} = {\bf L} \cdot
{\bf \Omega} = \frac{1}{2}{\bf L}{\cdot}({\bm
\nabla}{\times}\dot{\bf u})\, .
\end{equation}
While this equation has been derived for ${\bf L}$ of orbital
nature, it must equally apply to a spin. One comes to this
conclusion by considering the effect of the rotation in a
stationary coordinate frame which axes are determined by the local
crystal field at the location of the spin. The rotation of that
frame due to a transversal phonon is equivalent to the magnetic
field, which results in the same effective interaction with the
spin as it is for the orbital moment.

The other interactions of ${\bf L}$ with the phonon field are of
the magnetostriction form,
\begin{equation}\label{interactions}
{\cal{H}}_{int} = {\alpha}_{iklm}L_{i}L_{k}u_{lm}({\bf R}) + ...
\end{equation}
In the absence of the external field, the symmetry with respect to
the time reversal requires that all terms in Eq.\
(\ref{interactions}) contain even number of $L$-components. The
external field, ${\bf H}$, adds the Zeeman term, $-{\gamma}{\bf
L}{\cdot}{\bf H}$ to the Hamiltonian, with $\gamma$ being the
gyromagnetic ratio. Thus, the total Hamiltonian becomes
\begin{eqnarray}\label{total}
{\cal{H}} & = & {\beta}_{ik}L_{i}L_{k} +
{\beta}'_{iklm}L_iL_kL_lL_m + ... -{\gamma}{\bf L}{\cdot}{\bf H}
\nonumber \\
& + & \frac{1}{2}{\bf L}{\cdot}[{\bm \nabla}{\times}\dot{\bf
u}({\bf
R})] + {\alpha}_{iklm}L_{i}L_{k}u_{lm}({\bf R}) + ... \nonumber \\
& + & \int d^{3}r \left(\frac{1}{2}\rho {\dot{\bf u}}^{2} +
{\lambda}_{iklm}u_{ik}u_{lm}\right) \,.
\end{eqnarray}

The main difference of the quantum problem for ${\bf L}$ from the
quantum problem for ${\bf R}$ is that different components of the
operator $\hat{\bf L}$ do not commute with each other. For
certainty, we shall study the situation when the classical energy
minima of the potential correspond to the classical vector ${\bf
L}$ looking in one of the two directions along the $Z$-axis. This
could be, e.g., the case of a biaxial crystal field and a weak
magnetic field applied along the $Z$-axis, ${\cal{H}}_{0} =
-{\beta}_{z}L_{z}^{2}+{\beta}_{x}L_{x}^{2} -{\gamma}L_zH_z$, with
${\beta}_{z}, {\beta}_{x} > 0$. At ${\beta}_{x}=0$ the $| L
\rangle$ and $| -L \rangle$ eigenstates of ${\hat{L}}_z$, would
coincide with the eigenstates of ${\cal{H}}_0$. Since
${\hat{L}}_z$ does not commute with ${\hat{L}}_x$, any small
${\beta}_{x}L_{x}^{2}$ term mixes these states:
\begin{eqnarray}\label{L-states}
|0 \rangle & = & \frac{1}{\sqrt{2}}(C_{-}|L \rangle +
C_{+}|-L \rangle) \nonumber \\
|1 \rangle & = & \frac{1}{\sqrt{2}}(C_{+}|L \rangle - C_{-}|-L
\rangle)\;,
\end{eqnarray}
where $C_{\pm}$ are given by Eq.\ (\ref{C}) with $\epsilon =
2{\gamma}LH$.

As in the particle problem, the effect of the interactions in Eq.\
(\ref{total}) on the states $|0 \rangle$ and $|1 \rangle$ of Eq.\
(\ref{L-states}) is two-fold. Firstly, interactions renormalize
$\Delta$ in the expression for the energy gap, Eq.\ (\ref{gap}).
Secondly, they generate the finite width of $|1 \rangle$ due to
the finite probability of the decay $|1 \rangle \rightarrow |0
\rangle$. The latter, as in the particle problem, must be
considered with account of the symmetry of Eq.\ (\ref{total}). In
the quantum problem the products of the $L$-components in Eq.\
(\ref{total}) must be replaced by the symmetric combinations of
the ${\hat{L}}_i$-operators in order to preserve the Hermitian
property of the Hamiltonian, e.g., $L_iL_k \rightarrow
\frac{1}{2}({\hat{L}}_i{\hat{L}}_k + {\hat{L}}_k{\hat{L}}_i)$, or,
equivalently, the symmetry of the tensors ${\beta}_{ik}$, etc.
with respect to the transposition of indices must be enforced. It
is then easy to see that $\langle 0|({\hat{L}}_i{\hat{L}}_k +
{\hat{L}}_k{\hat{L}}_i)| 1 \rangle = 0$. This is true for any
combination of the even number of the operators ${\hat{L}}_i$.
Thus, the only term in Eq.\ (\ref{total}) responsible for the
decoherence of quantum oscillations between $L$ and $-L$ is
$\frac{1}{2}{\bf L}{\cdot}[{\bm \nabla}{\times}\dot{\bf u}({\bf
R})]$, which is independent of any interaction constants.
Consequently, the relevant matrix element of the operator of the
angular momentum is
\begin{equation}\label{L-element}
\langle 0| {\hat{L}}_z | 1 \rangle = LC_{+}C_{-} =
L{\omega}_0/{\omega} \,,
\end{equation}
and
\begin{eqnarray}\label{L-Fermi}
\Gamma  & = & \frac{\pi}{2{\hbar}^{2}}|\langle 0 | {\hat{L}}_z
  | 1 \rangle |^{2}  \nonumber \\
  & \times & \langle \sum_{{\bf k},i} |\langle n_{{\bf k}i}+1 |
  ({\bm \nabla}{\times}\dot{\bf u})_z| n_{{\bf k}i} \rangle |^{2}
\delta({\omega}_{{\bf k}i} - {\omega})\rangle_T \, ,
\end{eqnarray}
where $\langle ... \rangle_T$ means thermal average. Substituting
here the quantized phonon field of Eq.\ (\ref{u}), one obtains
\begin{equation}\label{L-T}
\Gamma  = \frac{L^{2}{\omega}_{0}^{2}{\omega}^{3}}{12 \pi {\hbar}
\rho{c_t^5}} \coth\left(\frac{{\hbar}{\omega}}{2T}\right) \,,
\end{equation}
which is the angular equivalent of Eq.\ (\ref{T}). This formula
generalizes the result of Ref. \onlinecite{CM} obtained for $H=0$.
Same as Eq.\ (\ref{T}), it contains no unknown interaction
constants.

As in the coordinate tunneling problem, it is interesting to
notice that at $H=0$ and $T=0$, Eq.\ (\ref{L-T}), can be written
as ${\hbar}{\Gamma} \sim L^{2}/2I$, where $I \sim {\rho}
{\lambda}^{5}$ is the moment of inertia of the part of the solid
of dimensions ${\lambda} {\sim} 2{\pi}c_t/{\omega}_0$, adjacent to
the particle (molecule) whose angular momentum tunnels between $L$
and $-L$. Thus, the physical origin of the decoherence given by
Eq.\ (\ref{L-T}) is the entanglement of the angular states of the
particle (molecule) with the angular states of the solid, required
by the conservation of the angular momentum.

In conclusion, we have studied quantum oscillations in a double
well coupled with a solid. When the oscillation frequency is small
compared to the Debye frequency, the decoherence due to phonons is
given by the universal formula which contains only directly
measurable parameters. It provides the ultimate lower limit for
the decoherence rate, mandated by the invariance with respect to
global translations and rotations.

This work has been supported by the NSF Grant No. EIA-0310517.
%



\end{document}